\documentclass[aps,prb,superscriptaddress,twocolumn]{revtex4-1}

\usepackage{setspace}

\usepackage{todonotes}

\usepackage{times}
\usepackage{amsmath}
\usepackage{amsfonts}
\usepackage{amssymb}
\usepackage{epsfig}
\usepackage{siunitx}
\usepackage{gensymb}

\usepackage{float}
\usepackage{ifthen}
\usepackage{xspace}

\usepackage{relsize}

\begin{document}


\title{Fractional statistics in anyon collisions}

\author{
H. Bartolomei$^{1 ,\dag}$, M. Kumar$^{1, \dag, \bot}$,  R. Bisognin$^{1}$, A. Marguerite$^{1,\ddag} $, J.-M. Berroir$^{1 }$, \\
 E. Bocquillon$^{1 }$, B. Pla\c{c}ais$^{1 }$,   A. Cavanna$^{2}$, Q. Dong$^{2 }$, U. Gennser$^{2 }$, Y. Jin$^{2 }$, and G. F\`{e}ve$^{1 \ast}$ \\
\normalsize{$^{1}$Laboratoire
de Physique de l\textquoteright Ecole normale sup\'erieure, ENS, Universit\'e
PSL, }\\
\normalsize{CNRS, Sorbonne Universit\'e, Universit\'e Paris-Diderot, Sorbonne Paris
Cit\'e, Paris, France}\\
\normalsize{$^{2}$ Centre de Nanosciences et de Nanotechnologies (C2N), CNRS, Univ. Paris Sud,}\\
 \normalsize{Universit\'{e} Paris-Saclay, 91120 Palaiseau, France.}\\
\normalsize{$^\ast$ To whom correspondence should be addressed; E-mail:  gwendal.feve@ens.fr.}\\
\normalsize{$^\dag$ These
authors contributed equally.}\\
\normalsize{$^\ddag$Present address: Department of Condensed Matter Physics,}\\
 \normalsize{Weizmann Institute of Science, Rehovot, Israel.}\\
\normalsize{$^\bot$Present address: Low Temperature Laboratory, Department of Applied Physics,}\\
\normalsize{ Aalto University, Espoo, Finland.}}

\date{}




\begin{abstract}
Two-dimensional systems can host exotic particles called anyons whose quantum statistics are neither bosonic nor fermionic. For example, the elementary excitations of the fractional quantum Hall effect at filling factor $\nu=1/m$ (where m is an odd integer) have been predicted to obey abelian fractional statistics, with a phase $\varphi$ associated  with the exchange of two particles equal to $\pi/m$. However, despite numerous experimental attempts, clear signatures of fractional statistics remain elusive. Here we experimentally demonstrate abelian fractional statistics at filling factor $\nu=1/3$ by measuring the current correlations resulting from the collision between anyons at a beam-splitter. By analyzing their dependence on the anyon current impinging on the splitter and comparing with recent theoretical models, we extract $\varphi=\pi/3$, in agreement with predictions.
\end{abstract}

\maketitle

In three dimensional space, elementary excitations fall into two categories depending on the phase $\varphi$ accumulated by the many-body wavefunction while exchanging two particles. This phase governs the statistics of an ensemble of particles: bosonic particles, for which $\varphi=0$, tend to bunch together whereas fermions ($\varphi=\pi$) antibunch and follow Pauli's exclusion principle. In two-dimensional systems, other values of $\varphi$ can be realized \cite{Leinaas1977,Wilczek1982}, defining  types of elementary excitations called anyons \cite{Wilczek1982b} that obey fractional or anyonic statistics with intermediate levels of bunching or exclusion. The fractional quantum Hall effect \cite{Tsui1982,Laughlin83}, obtained by applying a strong magnetic field perpendicular to a two-dimensional electron gas, is one of the physical systems predicted to host anyons. For a filling $\nu$ of the first Landau level belonging to the Laughlin series \cite{Laughlin83}, i.e. $\nu=1/m$ where $m$ is an odd integer, the exchange phase is predicted to be given by\cite{Halperin84,Arovas84} $\varphi=\pi/m$ interpolating between the bosonic and fermionic limits.
In spite of its fundamental interest, direct experimental evidences of fractional statistics remain elusive. To date, most efforts have focused on the implementation of single-particle interferometers \cite{Chamon97,Law2006} where the output current is expected to be directly sensitive to the exchange phase $\varphi$. However, despite many experimental attempts \cite{Camino2007,Ofek2010,McClure2012,Willett2013,Nakamura2019,Willet2019}, clear signatures are still lacking because the observed modulations of the current result not only from the variation of the exchange phase, but also from Coulomb blockade and Aharonov-Bohm interference \cite{Halperin2011}. In the case of non-abelian anyons \cite{Moore91}, where the exchange of quasiparticles is described by topological unitary transformations, recent heat conduction measurements showed evidence of a non-abelian state \cite{Banerjee,Kasahara}, though these results give only indirect evidence of the underlying quantum statistics.
 Here we follow a different approach and experimentally demonstrate that the elementary excitations of the fractional quantum Hall effect at filling factor $\nu=1/3$ obey fractional statistics with $\varphi=\pi/3$; we accomplish this by measuring the fluctuations or noise of the electrical current generated by the collision of anyons on a beam-splitter \cite{Rosenow2016}. The measurement of the current noise generated by a single scatterer of fractional quasiparticles \cite{dePicciotto1997,Saminadayar1997} has already successfully demonstrated that they carry a fractional charge $e^*=e/3$. Shortly after these seminal works, it was theoretically predicted \cite{Safi2001,Martin2005,Kim2005,Rosenow2016,Lee2019} that in conductors comprising several scatterers, noise measurements would exhibit two-particle interference effects where exchange statistics plays a central role, and would thus be sensitive to the exchange phase $\varphi$. In this context, current-current correlation measurements in collider geometries are of particular interest, as they have been extensively used to probe the quantum statistics of  particles colliding on a beam-splitter. In a seminal two-particle collision experiment, Hong-Ou and Mandel\cite{HOM} demonstrated that photons tend to bunch together in the same splitter output, as expected from their bosonic statistics. In contrast, collision experiments implemented in quantum conductors \cite{Liu1998,Ol'khovskaya2008,Bocquillon2013} have shown a suppression of the cross-correlations between the output current fluctuations caused by the antibunching of electrons, which are fermions.  This behavior can also be understood as a consequence of the Pauli exclusion principle that forbids two fermions to occupy the same quantum state at the splitter output. This exclusion principle can be generalized to fractional statistics \cite{Haldane1991,Isakov1999} by introducing an exclusion quasiprobability \cite{Rosenow2016} $p$ interpolating between the fermionic and bosonic limits. In a classical description of a two-particle collision, see Fig. 1A and [\onlinecite{Supp}], $p$ accounts for the effects of quantum statistics on the probability $K$ of finding two quasiparticles in the same output arm of the beam-splitter: $K=T(1-T)(1-p)$, where $T$ is the single particle backscattering probability (see Fig. 1A). The fermionic case is $p=1$, leading to perfect antibunching, $K=0$. Contrary to fermions, the bunching of bosons enhances $K$, meaning that $1-p>1$ and $p<0$.

 \begin{figure}[h!]
	\includegraphics[width=1
\columnwidth,keepaspectratio]{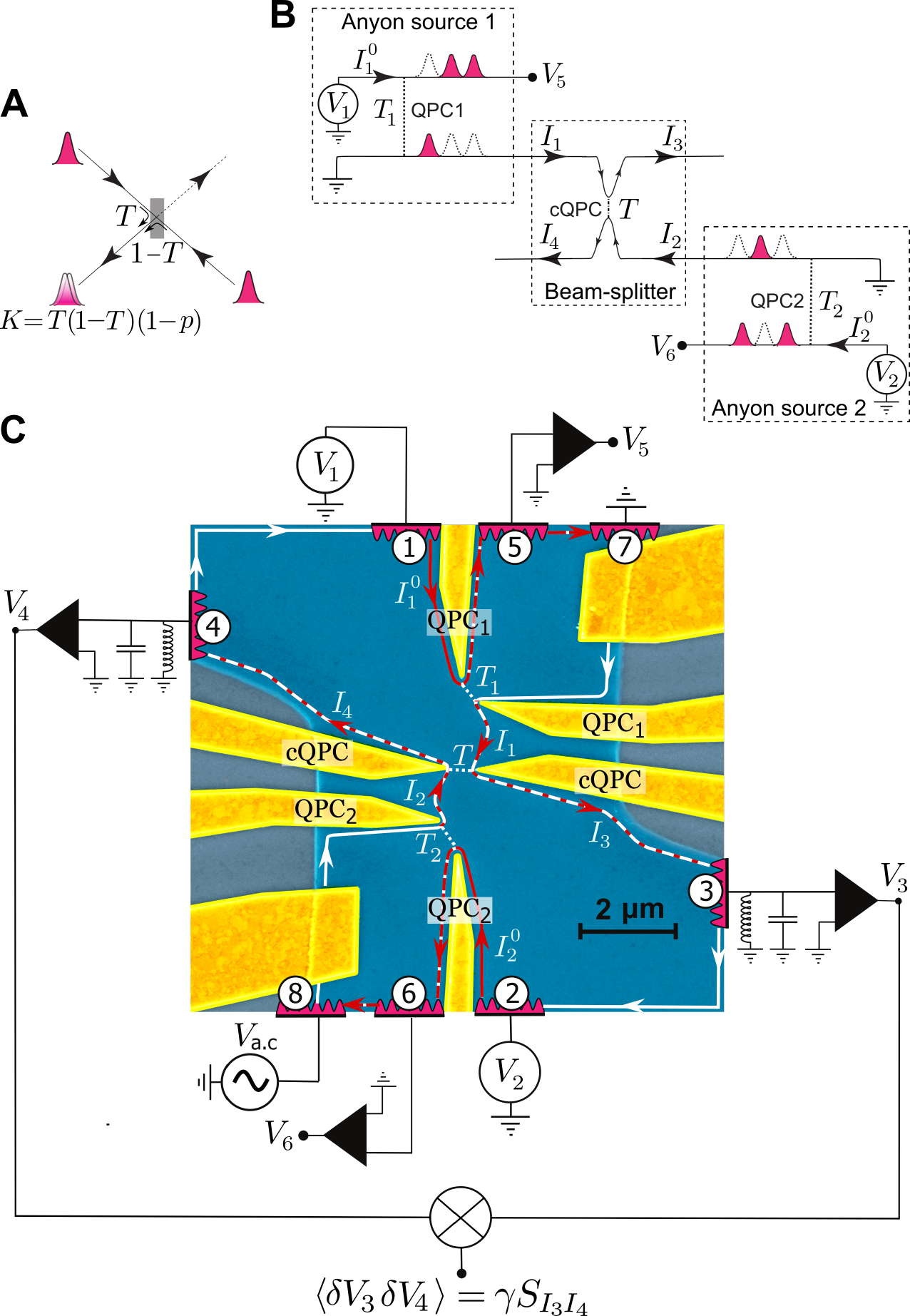}
	\caption{\label{fig1} \textbf{Figure1: sample and principle of the experiment.} (\textbf{A}) Exclusion quasiprobability $p$ : The probability $K$ to have two anyons exiting in the same output edge channel is modified by the factor $(1-p)$. (\textbf{B}) Principle of the experiment: The voltage $V$ generates the currents $I^{0}$ towards QPC1 and QPC2. These two QPC's tuned in the weak-backscattering regime $T_1, T_2\ll1$ act as random Poissonian sources of anyons which collide on the central QPC. (\textbf{C}) False colored scanning electron microscope (SEM) picture of the sample. The electron gas is represented in blue and the gates in gold. Edge currents are represented as red lines (red dashed lines after partitioning). }
\end{figure}

In order to implement collision experiments in quantum conductors, it is necessary to combine a beam-splitter for quasiparticles, a way to guide them ballistically, and two sources to emit them. The two first ingredients can be easily implemented in two-dimensional electron gases in the quantum Hall regime. Quantum point contacts (QPC) can be used as tunable beam-splitters and, at high magnetic field, charge transport is guided along the chiral edge channels. By combining these elements, single \cite{Ji2003} and two-particle \cite{Neder2007} electronic interferometers have been realized, and fermionic antibunching resulting from the collision between two indistinguishable electrons has been observed \cite{Bocquillon2013}. Investigating the anyonic case requires replacing the conventional electron sources (such as biased ohmic contacts) by sources of fractional anyonic quasiparticles. As suggested in Ref.[\onlinecite{Rosenow2016}], and as sketched in Fig. 1B, this implies using three QPC's. Two input QPC's labeled QPC1 and QPC2 are biased by dc voltages $V_1$ and $V_2$ and tuned in the weak backscattering regime to generate diluted beams of fractional quasiparticles. Indeed, it is known that, in the fractional quantum Hall regime, the partitioning of a dc electrical current $I^0$ with a small backscattering probability $T \ll 1$ occurs through the random transfer of quasiparticles of fractional charge\cite{Martin2005} $q=e^*$. As experimentally observed, the proportionality of the current noise \cite{dePicciotto1997,Saminadayar1997} with the input current $I^0$, the transmission $T$ and the fractional charge $e^*$, shows that this random transfer follows a Poissonian law. QPC1 and QPC2 can thus be used as Poissonian sources of anyons, which then collide on a third quantum point contact labeled cQPC; cQPC is used as a beam-splitter in the collision experiment. The fractional statistics of the colliding quasiparticles can be revealed by measuring the cross-correlations between the electrical currents at the output of the beam-splitter.
The sample (Fig. 1C) is a two-dimensional electron gas (GaAs/AlGaAs). The magnetic field is set to $B=13$ T, corresponding to a filling factor $\nu=1/3$ for a charge density $n_s=1.09 \times 10^{15}$ m $^{-2}$. At this field and at very low electronic temperature $T_{\text{el}}=30$ mK,  ballistic charge transport occurs along the edges of the sample without backscattering (see [\onlinecite{Supp}] for more details). As discussed above, the two quasiparticle sources comprise two quantum point contacts with transmissions $T_1$ and $T_2$ ($T_1$, $T_2 \ll 1$). We apply the voltages $V_1$ and $V_2$ to ohmic contacts 1 and 2 in order to generate the input currents $I_{1/2}^0=\frac{e^2}{3h}  V_{1/2}$ (where $h$ is the Planck's constant) towards QPC1 and QPC2. They randomly generate the quasiparticle currents $I_1$ and $I_2$ propagating towards the third quantum contact cQPC, which is used as a beam splitter with transmission $T$. $I_1$ and $I_2$ are extracted from the measurement of the voltages at output contacts 5 and 6 with $V_{5/6}=\frac{3h}{e^2} (I_{1/2}^0-I_{1/2})$. The transmission $T$ is extracted from the measurement of the small ac current $I_{\text{ac}}$ flowing in contact 3 when a small ac voltage $V_{\text{ac}}$ is applied on contact 8: $T=\frac{3h}{e^2}   \frac{1}{1-T_2} \frac{ I_{\text{ac}}}{V_{\text{ac}}}$ . We measure the low frequency correlations between the current fluctuations at the splitter outputs $\delta I_3$ and $\delta I_4$, defined as $S_{I_3 I_4 }=2 \int d\tau \langle \delta I_3 (t) \delta I_4 (t+\tau) \rangle$, with $\delta I_3 (t)=I_3 (t)-\langle I_3 (t)\rangle$. They are measured as voltage fluctuations across the quantized Hall resistance $R_K=3h/e^2$ , which converts the currents $I_{3/4}$ into the voltages $V_{3/4}=\frac{3h}{e^2} I_{3/4}$. $V_3$ and $V_4$ are then amplified by a combination of homemade cryogenic and room temperature low noise amplifiers. The output voltage cross-correlations, $\langle \delta V_3 \delta V_4 \rangle$, are then measured by integrating $S_{V_3 V_4 }$ in a bandwidth $\delta f$ centered on the frequency $f_0=1.11$ MHz, set by the LC tank circuits connected to contacts 3 and 4. The bandwidth $\delta f=20$ kHz is set by the quality factor, $Q\approx 55$, of the LC resonators in parallel with $R_K$. The output voltage cross-correlations are directly proportional to the input current correlations $\langle \delta V_3 \delta V_4 \rangle =\gamma S_{I_3 I_4 }$, see [\onlinecite{Supp}] for the calibration of the conversion factor $\gamma$.
We focus first on the case $I_1=I_2$, by setting the two input QPC's to equal transmission $T_S=T_1=T_2$, and by applying the same voltage $V_1=V_2$ at the input ohmic contacts 1 and 2. Defining the total input current $I_+=I_1+I_2$ and the current difference $I_-=I_1-I_2$, this setting corresponds to a vanishing current difference between the splitter input arms, $I_-=0$. In this regime, the differences between fermionic and anyonic statistics are emphasized. Indeed, a fermionic behavior ($p=1$) results in a full suppression of the output cross-correlations $S_{I_3 I_4 }=0$. In contrast, for anyons obeying fractional statistics ($p \neq 1$) negative cross-correlation $S_{I_3 I_4} <0$ is expected.  In the classical description \cite{Rosenow2016,Supp} and in the case $I_-=0$, $S_{I_3 I_4,cl}$ is directly proportional to the probability $K$, and hence contains the information on the exclusion quasiprobability $p$:
\begin{equation}
	S_{I_3 I_4,cl}=-2e^* (1-p)T_S T(1-T)I_+	
\end{equation}
 Equation 1 shows that $S_{I_3 I_4 }$ is directly proportional to the total input current $I_+$, allowing us to define a generalized Fano factor $P=S_{I_3 I_4 }/[2e^* T(1-T)I_+ ]$. The classical prediction is $P=-(1-p)T_S$ showing that $P$ is the relevant parameter carrying the information on statistics. The classical calculation thus provides valuable insights into the connection between the measurement of current cross-correlations in a collision experiment and quantum statistics. However, the accurate prediction of $S_{I_3 I_4 }$ requires a complete quantum mechanical description of anyon collisions, where the nonequilibrium dynamics of the chiral edge channels are described by bosonic fields \cite{Levkivskyi2016}. Within this framework, the current cross-correlations, resulting from the collisions between anyons randomly emitted by Poissonian sources, are predicted to directly reflect the braiding statistics of anyons. In the case $I_-=0$ and for anyons with an exchange phase $\varphi=\pi/m$ (with $m\geq3$), the cross-correlations $S_{I_3 I_4,q}$ are predicted to vary linearly with the total input current $I_+$, with a generalized Fano factor\cite{Rosenow2016} explicitly related to $\varphi$:
\begin{equation}
	S_{I_3 I_4,q}=2e^*  \frac{-2}{m-2} TI_+   \quad  (T\ll 1)	
\end{equation}
 Equation 2 shows that the generalized Fano factor $P$ is directly related to the exchange phase, $P=-2/(m-2)$ and is independent of $T_S$ (in the limit $T_1=T_2=T_S \ll 1$). In particular, for the filling factor $\nu=1/3$ ($\varphi=\pi/3$), the prediction is $P=-2$, which strongly differs from the fermionic behavior ($P=0$). In the light of the classical model (Equation 1) the prediction $P=-2$ suggests $p<0$ corresponding to a bunching behavior which is expected at $\nu=1/3$ as the exchange phase\cite{Halperin84,Arovas84}, $\varphi=\pi/3$, is closer to the bosonic value ($\varphi=0$) than the fermionic one ($\varphi=\pi$). Our main result is the experimental measurement of $P=-2$ at $\nu=1/3$, providing an experimental demonstration of anyonic statistics with an exchange phase $\varphi=\pi/3$ in agreement with predictions.

 \begin{figure}[h!]
	\includegraphics[width=1
\columnwidth,keepaspectratio]{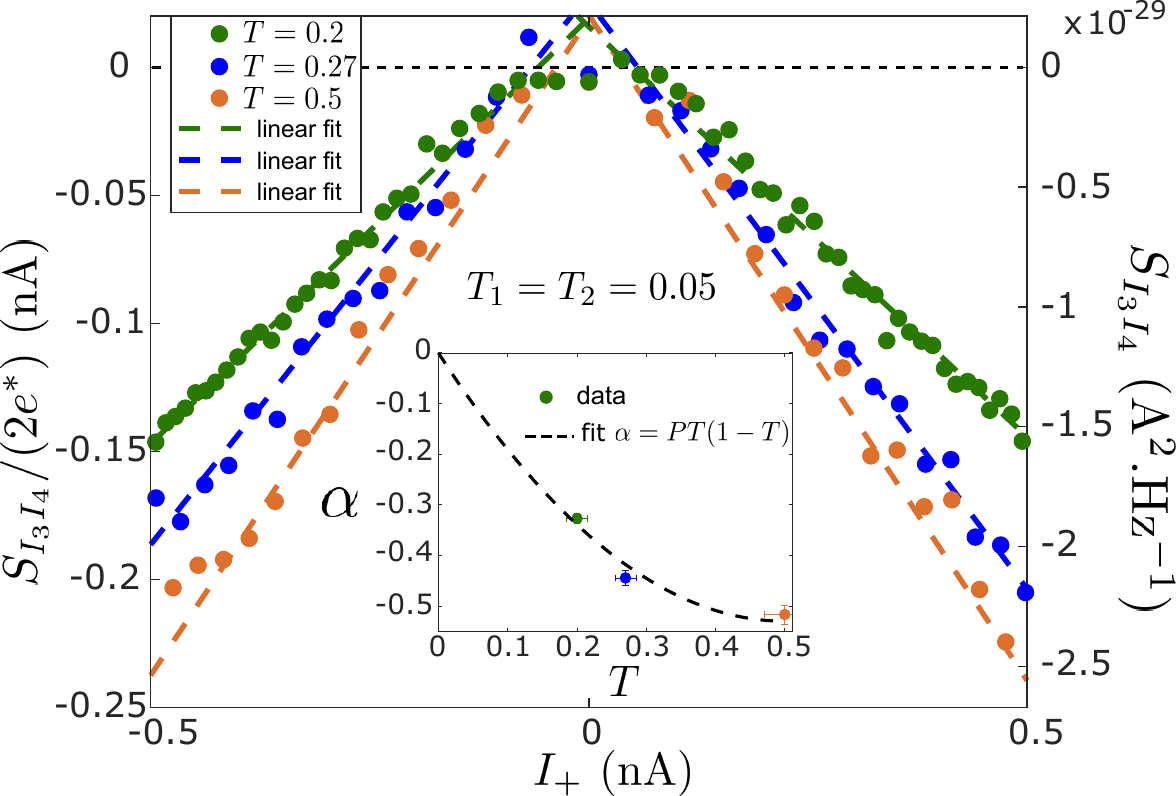}
	\caption{\label{fig2} \textbf{Figure 2: Fano factor in anyon collision}. $S_{I_3 I_4}$ for $T_1=T_2=0.05$ as a function of $I_+$ and for various transmissions $T$ of the central QPC. The dashed lines are linear fits of $S_{I_3 I_4}/(2e^{*})$. Inset, slope $\alpha$ extracted from the linear fits as a function of the central QPC transmission $T$. The dashed line is a fit to $\alpha=PT(1-T)$ with $P=-2.1 \pm 0.1$. }
\end{figure}

In order to enforce the Poissonian emission of fractional quasiparticles from QPC1 and QPC2, we set $T_S=0.05$, in the weak backscattering regime. We then measure the cross-correlations $S_{I_3 I_4 }$ resulting from the quasiparticle collisions as a function of the total current $I_+$ for different transmissions $T$ of the beam-splitter\cite{Note}, $T=0.2$, $T=0.27$ and $T=0.5$, see Fig. 2. In the three cases, for $I_+\geq 50$ pA, negative cross-correlations varying linearly with the current $I_+$ are observed. We extract the slope $\alpha$ of the variation of $S_{I_3 I_4 }/(2e^*)$ by a linear fit (dashed lines) of the experimental data. The three extracted values of $\alpha$ are plotted in the inset  to Fig. 2 as a function of the beam-splitter transmission $T$. The observed $T$ dependence agrees with the binomial law $T(1-T)$ (dashed line) extending Eq. 2 for transmissions beyond the weak-backscattering regime \cite{Isakov1999}. The generalized Fano factor can be extracted from the fit of $\alpha$ with the dependence $\alpha=PT(1-T)$, giving $P=-2.1 \pm 0.1$, and demonstrating the fractional statistics at $\nu=1/3$ with the predicted exchange phase $\varphi=\pi/3$. In striking contrast, we observe $P\approx 0$ at filling factor $\nu=2$, see Fig. S5, corresponding to the expected fermionic behavior for integer filling factor.

\begin{figure}[h!]
	\includegraphics[width=1
\columnwidth,keepaspectratio]{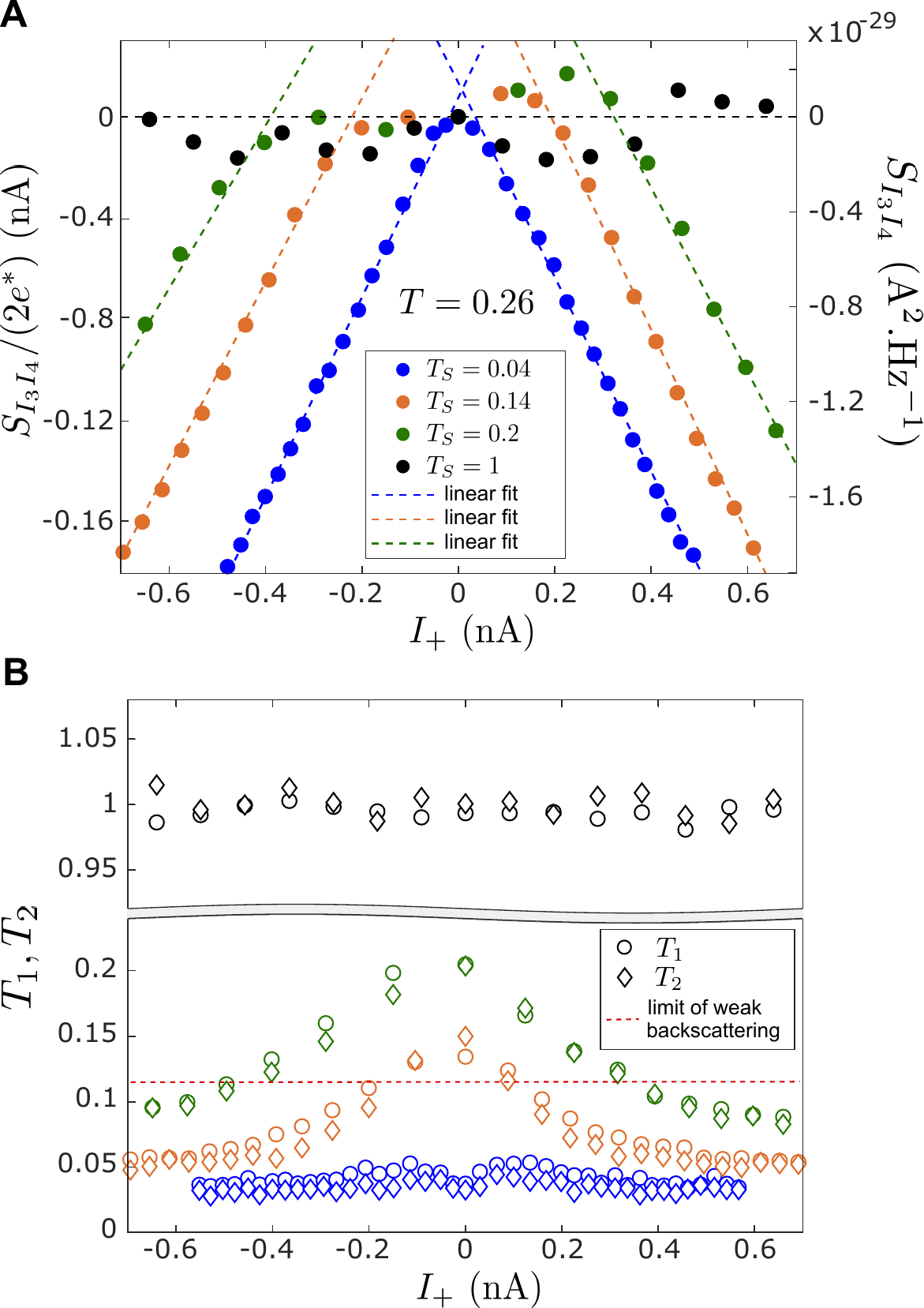}
	\caption{\label{fig3} \textbf{Figure 3: anyonic vs fermionic behavior.} (\textbf{A}) $S_{I_3 I_4}$ as a function of $I_+$ for various transmissions of the input QPCs (measured at $I_+=0$): $T_S=0.04$,  $T_S=0.14$, $T_S=0.2$, and $T_S=1$. The dashed lines are linear fits of $S_{I_3 I_4}$. (\textbf{B}) Transmission $T_1$ and $T_2$ as a function of $I_+$. The red dashed line sets the limit of the weak-backscattering regime where the anyonic behavior $P\approx -2$ is observed. }
\end{figure}

The fermionic behavior can be restored at $\nu=1/3$ by increasing the transmissions of the input QPC's $T_1$ and $T_2$, thereby deviating from the weak-backscattering regime suitable for the emission of anyons. For $T_S=1$ (black points in Fig. 3A), we observe fermionic behavior: $S_{I_3 I_4 }=0$ for all values of $I_+$. For intermediate values of $T_S$, the I-V characteristics of the input QPC's are strongly non-linear (see Fig. 3B). $T_S$ decreases when $I_+$ is increased, eventually restoring the weak-backscattering limit at large bias. The measurements of $S_{I_3 I_4 }$ for $T_S=0.14$ and $T_S=0.2$ (for $I_+=0$) plotted on Fig. 3A reflect this evolution. At low current $I_+$, fermionic behavior is observed, $S_{I_3 I_4}=0$. At higher current, where the weak-backscattering limit is restored, the linear evolution of the cross-correlations with $I_+$ is recovered, with a generalized Fano factor almost constant. $P$ slightly increases from $P=-2.00 \pm 0.15$ for $T_S=0.04$ to $P=-1.94 \pm 0.12$ for $T_S=0.14$, and $P=-1.73 \pm 0.10$ for $T_S=0.2$. As expected, the domain where the fermionic behavior is observed ($S_{I_3 I_4 }=0$) increases when the transmission $T_S$ increases; it varies from $|I_+ |\leq 200$ pA at $T_S=0.14$ to $|I_+ | \leq 400$ pA at $T_S=0.2$.  These data confirm that $P=-2$ is observed only in the regime of anyon emission, and that regular fermionic behavior $P\approx 0$ takes place away from the weak-backscattering limit.

\begin{figure}[h!]
	\includegraphics[width=1
\columnwidth,keepaspectratio]{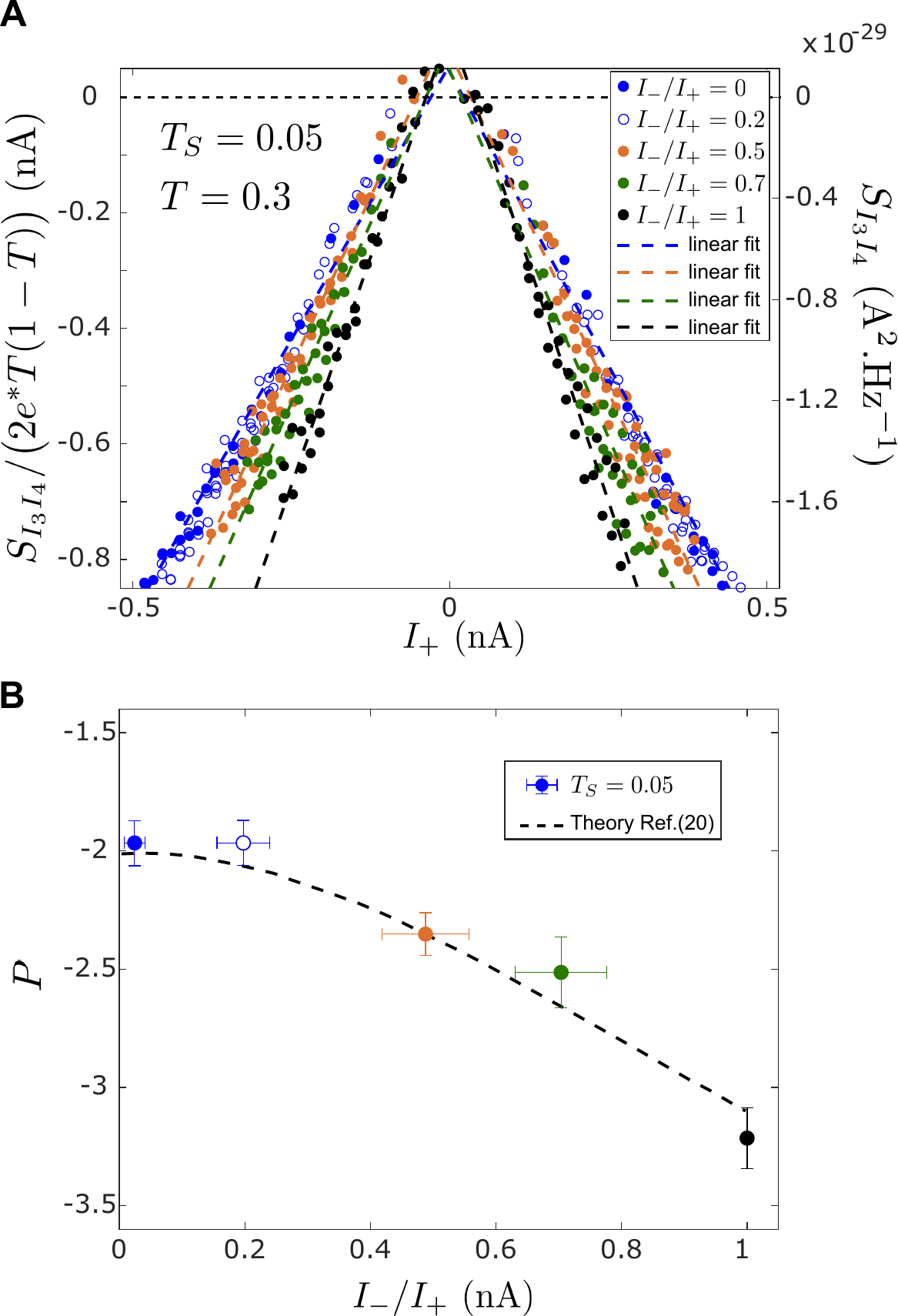}
	\caption{\label{fig4} \textbf{Figure 4: experimental test of the quantum mechanical description of an anyon collision.} (\textbf{A}) $S_{I_3 I_4}$ as a function of $I_+$ for various values of the ratio $I_-/I_+$. The dashed lines are linear fits of $S_{I_3 I_4}$.  (\textbf{B}) Generalized Fano factor $P$ extracted from the slope of the linear fits plotted as a function of the ratio $I_-/I_+$. The dashed line is the prediction extracted from Ref.[\onlinecite{Rosenow2016}] for the quantum description of anyon collisions with $\varphi=\pi/3$ .}
\end{figure}

We finally check in more detail the agreement between our measurements and the quantum description \cite{Rosenow2016} of anyon collisions, by investigating the dependence of the Fano factor P on the ratio $I_-/I_+$. Contrary to the previous experiments where $I_-=0$ was imposed by $V_1=V_2$ and $T_1=T_2$, we instead modify the ratio $I_-/I_+$ by varying the values of the input voltages $V_1 \neq V_2$. Figure 4A presents the evolution of $S_{I_3 I_4 }$ as a function of the total current $I_+$ for four different values of the ratio $I_-/I_+$ and $T_S=0.05$. We observe in the four cases a linear evolution with $I_+$, with a slope $P$ that decreases when $I_-/I_+$ increases. The different values of $P$ extracted from a linear fit of the data (dashed lines) are plotted on Fig. 4B. For $I_-/I_+ \leq 0.2$, $P$ is constant with $P\approx -2$. $P$ then decreases linearly towards $P \approx -3$ for $I_-/I_+ \approx 1$. These experimental results can be compared with the calculation of Ref.[\onlinecite{Rosenow2016}] (dashed line). The excellent agreement between our experimental results and the calculations further supports the quantum description of anyons with $\varphi=\pi/3$.
Our measurement of the Fano factor $P=-2$ demonstrates the anyonic statistics of the charge carriers with an exchange phase $\varphi=\pi/3$ in accordance with the predictions for the Laughlin state $\nu=1/3$. Interestingly, the prediction $P=-2$ for $\nu=1/3$ is valid when edge reconstruction effects can be neglected. Although neutral modes have been observed\cite{Inoue2014} even at $\nu=1/3$, the agreement with the prediction for a simple edge structure suggests that their effect can be neglected in our experiment, see [\onlinecite{Supp}]. Collision experiments similar to ours could be used to characterize the elementary excitations of other fractional quantum Hall phases with different fractional statistics or even more exotic cases where non-abelian statistics \cite{Moore91} are predicted. Additionally, combining collision experiments with the triggered emission of fractional quasiparticles \cite{Rech2017,Kapfer2018} would allow one to perform on-demand braiding of single anyons in a quantum conductor.

\section*{Acknowledgments}
 Funding: This work has been supported by the ANR grant "1shot reloaded" (ANR-14-CE32-0017), the ERC consolidator grant "EQuO" (No. 648236), and the French RENATECH network. Author Contributions: AM and GF designed the sample. YJ fabricated the sample on GaAs/AlGaAs heterostructures grown by AC and UG. YJ and QD designed and fabricated the low-frequency cryogenic amplifiers. MK, HB and RB conducted the measurements. HB, MK, RB, AM, JMB, EB, BP and GF participated in  the analysis and the writing of the manuscript with inputs from YJ and UG. GF supervised the project. Competing interests: The authors declare no competing interests. Data and materials availability: All data are available in \onlinecite{Zenodo}.\\
 
\textbf{Supplementary Materials}\\
https://science.sciencemag.org/content/368/6487/173/suppl/DC1 \\
Materials and Methods \\
Supplementary Text \\
Figs. S1 to S8 \\
References (42-44) \\


\begin{thebibliography}{9}
\bibitem{Note}
Due to a depletion of the central QPC by the gates at the center of the sample, $T=0.2$ is the lowest experimentally reachable value.
\end{thebibliography}
\end{document}